\documentclass[runningheads]{llncs}
\usepackage[T1]{fontenc}
\usepackage{graphicx}
\usepackage{color}

\usepackage{amsmath}
\usepackage{xcolor}
\usepackage{amssymb} 
\usepackage{booktabs}

\usepackage{multicol}
\usepackage{multirow}

\usepackage{pifont}

\begin{document}
\title{CCPL: Cross-modal Contrastive Protein Learning}
%
%

\author{Jiangbin Zheng\inst{1,2}\orcidID{0000-0003-3305-0103} \and
Stan Z. Li\inst{2,*}\orcidID{0000-0002-2961-8096}}
\authorrunning{J. Zheng et al.}
%
\institute{Zhejiang University, Hangzhou, China \and
AI Lab, Westlake University, Hangzhou, China  \\
* Corresponding Author
\\
\email{\{zhengjiangbin,Stan.ZQ.Li\}@westlake.edu.cn}}

\maketitle              
%
\begin{abstract}
Effective protein representation learning is crucial for predicting protein functions. Traditional methods often pretrain protein language models on large, unlabeled amino acid sequences, followed by finetuning on labeled data. While effective, these methods underutilize the potential of protein structures, which are vital for function determination. Common structural representation techniques rely heavily on annotated data, limiting their generalizability. Moreover, structural pretraining methods, similar to natural language pretraining, can distort actual protein structures. In this work, we introduce a novel unsupervised protein structure representation pretraining method, cross-modal contrastive protein learning (CCPL). CCPL leverages a robust protein language model and uses unsupervised contrastive alignment to enhance structure learning, incorporating self-supervised structural constraints to maintain intrinsic structural information. We evaluated our model across various benchmarks, demonstrating the framework's superiority.

\end{abstract}

\section{Introduction}
Learning effective protein representations is crucial for various biological tasks. In recent years, deep protein representation learning has revolutionized the field, notably in protein structure prediction, represented by AlphaFold2~\cite{jumper2021highly}, and protein design, exemplified by \cite{hsu2022learning} and ProteinMPNN \cite{dauparas2022robust}.
With the advent of low-cost sequencing technologies, a vast number of new protein sequences have been discovered. Current methods typically pretrain protein language models on large, unlabeled amino acid sequences\cite{rao2020transformer,lin2022language} and then finetuned on downstream tasks using limited labeled data. While sequence-based methods are effective, they often fail to explicitly capture and utilize existing protein structural information, which is vital for protein functions.

Given the high cost and time-consuming nature of annotating new protein functions, there is a pressing need for accurate and efficient function annotation methods to bridge the existing sequence-function gap. Since the functions are governed by folded structures, some data-driven approaches rely on learning structural representations of proteins, which are then applied to various tasks such as protein design, and function prediction and classification. Therefore, an effective structural encoder is essential. To better leverage structural information, several structure-based protein encoders have been proposed \cite{2021Intrinsic,hermosilla2022contrastive}. However, due to the scarcity of protein structures, these encoders are often designed for specific tasks, and their generalizability to other tasks remains unclear.

Protein structure encoders face two main challenges: 1) Data scarcity. The number of reported protein structures is significantly lower than datasets in other machine learning fields due to the challenges of experimental protein structure determination. For example, the Protein Data Bank (PDB) contains 182K experimentally determined structures, whereas Pfam has 47 million protein sequences \cite{mistry2021pfam} and ImageNet contains 10 million annotated images. 2) Representation difficulty. Unlike sequences, traditional self-supervised language pretraining methods, such as masked language modeling, are not feasible for learning structural representations. Introducing noise or perturbations into structural data can lead to unstable or chemically incorrect structures, making augmented data unreliable. Therefore, the ability to pretrain on known protein structures has not been widely applied to protein property prediction.

By rethinking the protein representations, we observe that the success of sequence-based models is due to large-scale data and the guidance provided by self-supervised signals. Considering there is a natural pairing relationship between structure and sequence, establishing this relationship can help guide structural learning without compromising the protein structure itself. Based on this observation, we pose the question: \textit{Can we augment protein structure model training supervised by robust pretrained protein language models?}

Inspired by advances in cross-modal pretraining (e.g., CLIP~\cite{ramesh2022hierarchical}, Context-to-Vector~\cite{zheng2022using}), we introduce a novel cross-modal contrastive protein learning (CCPL). This method calculates contrastive loss between two independently pretrained encoders, maximizing the similarity between paired protein structures and sequences while minimizing it for non-paired ones~\cite{zheng2023cvt,zheng2022leveraging,zheng2021enhancing,zheng2020improved}. Compared to supervised learning methods, our contrastive learning approach has several advantages. First, finding the matching relationship between protein structures and sequences is natural. Second, our designed contrastive loss reduces dependence on explicit functional annotations, facilitating the use of large-scale unlabeled data.
We reframe structural representation training as an information retrieval task, where the protein structure is the query, and the goal is to retrieve the sequence with the highest binding probability to the target protein structure from a pool of candidates. To strengthen structural representation constraints, we also propose a self-supervised contact map constraint based on intermediate features from the structural encoder.

\begin{figure*}[htp]
  \centering
   \includegraphics[width=0.86\linewidth]{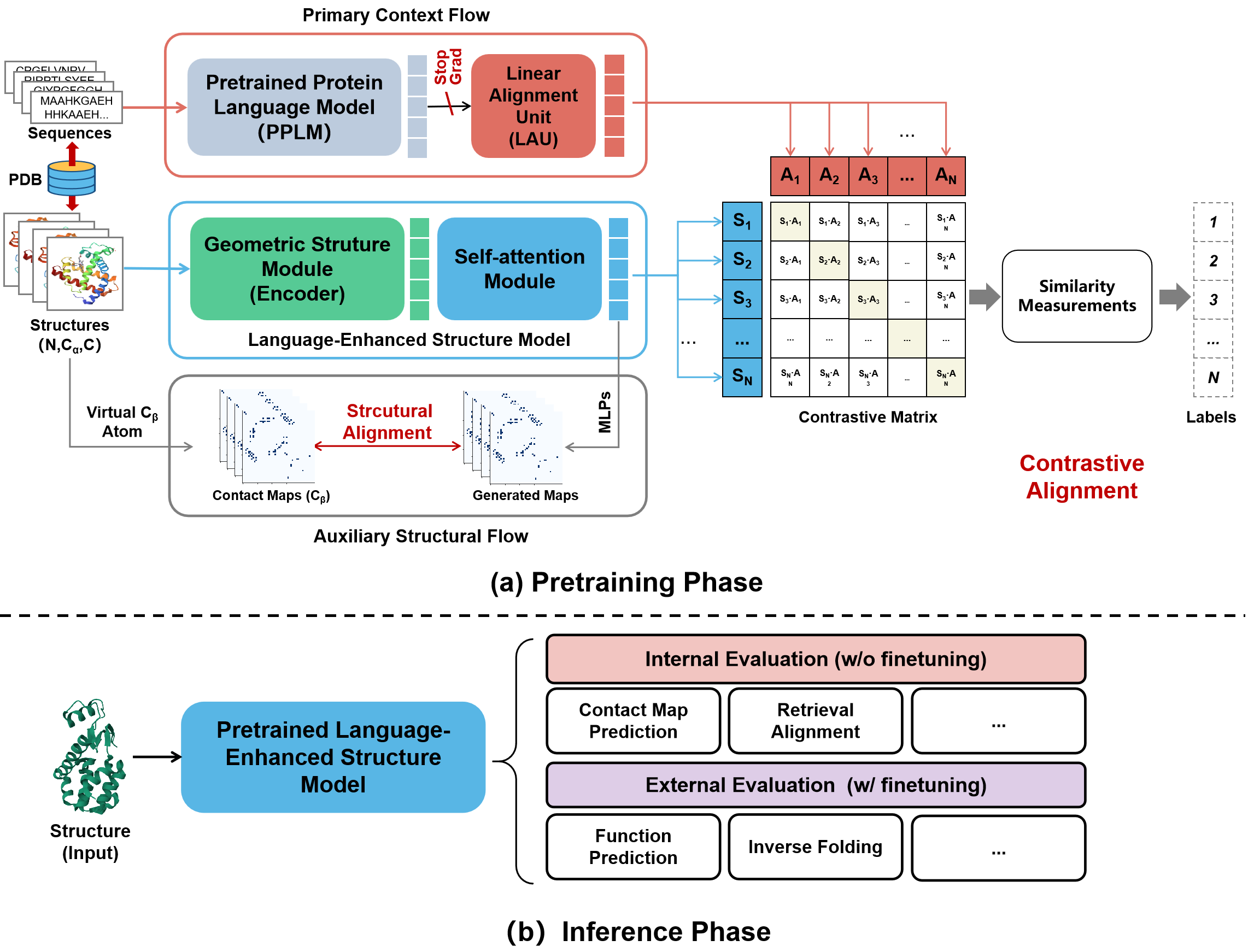} 
   \vspace{-1em}
   \caption{(a) The proposed cross-modal contrastive learning framework utilizes a pretrained protein language model to guide the training of the protein structure model through contrastive alignment loss. To reinforce information constraints on the structure, we introduce a self-supervised contact map prediction. (b) The internal and external evaluation tasks for our trained structure model during inference phase.}
   \label{fig:overall}
   \vspace{-1.2em}
\end{figure*}

To evaluate our proposed framework, we conducted benchmark tests. Due to the lack of established evaluation strategies for this novel pretraining paradigm, we designed a series of evaluation experiments, including internal tasks (e.g., contact map prediction and distribution alignment quality assessment) to demonstrate internal contrastive alignment ability and refinement performance, and external/downstream tasks (e.g., protein design and function prediction) to demonstrate generalization capability. Our experimental results validate the effectiveness of the CCPL framework, highlighting its robustness in pretraining performance and exceptional downstream task performance.

Our contributions can be summarized as follows:

$\bullet$ We propose an cross-modal protein representation framework, establishing a novel deep alignment relationship between sequences and structures. For the first time, we pretrain protein structural models under the guidance of rich prior language knowledge from pretrained protein sequence models.

$\bullet$ We introduce a comprehensive evaluation system to assess the pretrained structural models, providing benchmarks for the protein research community.

$\bullet$ Our proposed protein structural model pretraining demonstrates competitive performance across various evaluation tasks.

\section{Proposed CCPL Framework}

\subsection{Problem Statements}

\begin{figure*}[htp]
  \centering
   \includegraphics[width=0.7\linewidth]{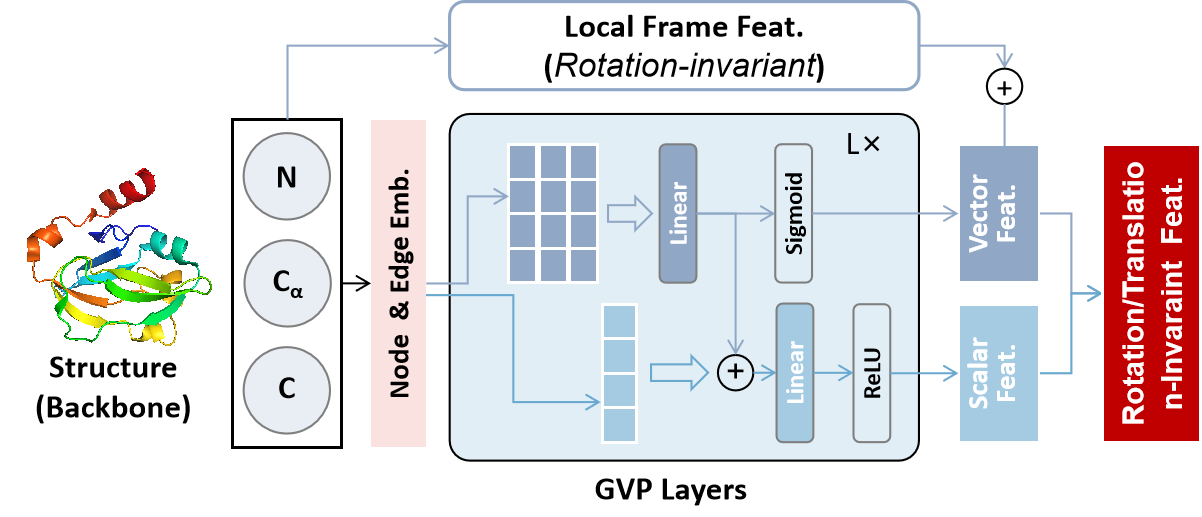}
   \vspace{-1em}
   \caption{Schematic diagram of the GVP module: Protein backbone atoms (C, C$_\alpha$, and N) form the basis for generating graphs with node and edge features based on k-nearest neighbor relationships. These graphs are fed into the vector and scalar channels of the GVP module to produce vector and scalar features. These primary features are then enhanced with additional spatial features, including rotation frame, sidechain, orientation, and dihedral characteristics, to create comprehensive spatial structure features.}
   \label{fig:gvp}
   \vspace{-1em}
\end{figure*}

Viewing CCPL as a supervised pseudo-dense retrieval task, we treat the protein structure as a query and retrieve the most relevant sequence from a given dataset. The overall framework, illustrated in Figure~\ref{fig:overall}, involves training two separate encoders to learn representations for protein structures and sequences, respectively. The similarity between each protein structure-sequence pair is then calculated, and a contrastive learning objective is used to distinguish between positive and negative pairs. 
Formally, given a protein structure \( p \) and its paired protein sequence \( m \), the objective of CCPL representation training is to maximize the probability of the sequence that naturally pairs with the structure. This selection process is guided by a scoring function \( \delta(\cdot) \), which evaluates the pairing probability between the protein \( p \) and the candidate sequence \( m \).

\subsection{Protein Structure and Sequence Encoders}

\textbf{Equivariant Protein Structure Encoder.}
For downstream protein structure tasks, it is essential that the predicted sequences remain unaffected by the reference frame of the structural coordinates. This means that the model's output distribution should be invariant under any rotation or translation applied to the input coordinates \( H \). Equivariant network models, such as geometric vector perceptron (GVP)-based models, are commonly employed in protein 3D structure modeling to meet this requirement. Models like GVP-GNN \cite{jing2020learning} and GVP-Transformer \cite{hsu2022learning} have demonstrated impressive performance in protein design tasks, highlighting the crucial role of the GVP module in representing structural features while maintaining equivariant properties.
Thus, we use GVP as our equivariant structure module. For any transformation \(\mathcal{T} \in E(3)\), the GVP module \(\varphi\) is considered E(3)-equivariant since \(\varphi(\mathcal{T} \cdot H) = \mathcal{T} \cdot \varphi(H)\), and E(3)-invariant since \(\varphi(\mathcal{T} \cdot H) = \varphi(H)\). Moreover, the simplicity and lightweight nature of GVP components make the GVP architecture well-suited for our structure module. A schematic diagram of the GVP module is illustrated in Figure~\ref{fig:gvp}.

\textbf{Pretrained Protein Language Encoder.}
Pretrained protein language models have become integral to deep protein downstream tasks, having been trained on extensive datasets. These sequence models inherently encapsulate rich structural information gleaned from sequence data, enabling them to effectively guide structure model learning. Among the available models, we choose ESM-2 \cite{lin2022language} as our primary teacher model due to its exceptional performance.

\textbf{Formulation.}
Formally, we denote the protein structure encoder as \( g_\phi \) with parameters \( \phi \) and the sequence encoder as \( f_\theta \) with parameters \( \theta \). The representations of the protein structure vector \( x^p \) and the corresponding sequence vector \( y^m \) are then defined as \( g_\phi(x^p) \) and \( f_\theta(y^m) \), respectively.

\subsection{Training Objectives}

\textbf{Contrastive Alignment Objective.}
To enable the contrastive learning process, we first need to measure the similarity between each protein structure and sequence pair. Drawing on previous research, we can utilize either dot product or cosine similarity for this purpose. When using the dot product, the similarity score for pairs \((x_{p}^{i}, y_{m}^{j})\), where \(i, j \in [1, N]\), is defined as:
\begin{equation}
\delta(x_{p}^{i}, x_{m}^{j}) = g_{\phi}(x_{p}^{i})^{T} \cdot f_{\theta}(y_{m}^{j}),
\end{equation}
where, by normalizing these scores, we then obtain cosine similarity.
Since our protein dataset includes only positive pairs of binding protein structures and sequences, we need to create negative pairs for contrastive learning. We implement a batch-wise sampling strategy inspired by CLIP. For a given batch of paired data \(\{(x^{p}_{b}, y^{m}_{b})\}_{b=1}^{B}\) with batch size \(B\), we extract a list of protein structures \(\{x^{p}_{b}\}_{b=1}^{B}\) and a corresponding list of sequences \(\{y^{m}_{b}\}_{b=1}^{B}\). By combining these lists, we generate \(B^2\) pairs \((x^{p}_{i}, y^{m}_{j})\) where \(i, j \in [1, B]\). Pairs where \(i = j\) are positive, while pairs where \(i \neq j\) are negative.
This approach relies on a fundamental assumption: if a protein and sequence pair is known to bind, it is likely that this protein does not bind with other sequences and vice versa. This assumption is validated by the distinct distribution patterns of positive and negative pairs.

We formalize this with two loss functions: the structure-to-sequence loss \( \mathcal{L}^{p} \) and the sequence-to-structure loss \( \mathcal{L}^{m} \) . The structure-to-sequence loss quantifies the likelihood of ranking the correct binding sequence higher than other sequences for a given protein structure \(x_{p}^{b}\):
\begin{equation}
\mathcal{L}_{b}^{p}(x_{b}^{p}, \{y_{i}^{m}\}_{i=1}^{B}) = -\frac{1}{B}\sum_{i=1}^{B} \log \frac{\exp(\delta(x_{b}^{p}, y_{i}^{m}))}{\sum_{j=1}^{B} \exp(\delta(x_{b}^{p}, y_{j}^{m}))}
\end{equation}

Conversely, the sequence-to-structure loss measures the likelihood of correctly ranking the binding target for a given molecule \(y_{m}^{b}\):
\begin{equation}
\mathcal{L}_{b}^{m}(y_{b}^{m}, \{x_{i}^{p}\}_{i=1}^{B}) = -\frac{1}{B}\sum_{i=1}^{B} \log \frac{\exp(\delta(y_{b}^{m}, x_{i}^{p}))}{\sum_{j=1}^{B} \exp(\delta(y_{b}^{m}, x_{j}^{p}))}
\end{equation}

Therefore, the final contrastive alignment loss for a mini-batch is the average of the two-direction losses $\mathcal{L}_{\text{align}}$ as:
\begin{equation}
    \mathcal{L}_{\text{align}} = \frac{1}{2} \sum_{b=1}^{B}(\mathcal{L}_b^p + \mathcal{L}_b^m) .
\end{equation}

\begin{figure*}[htp]
  \centering
   \includegraphics[width=0.95\linewidth]{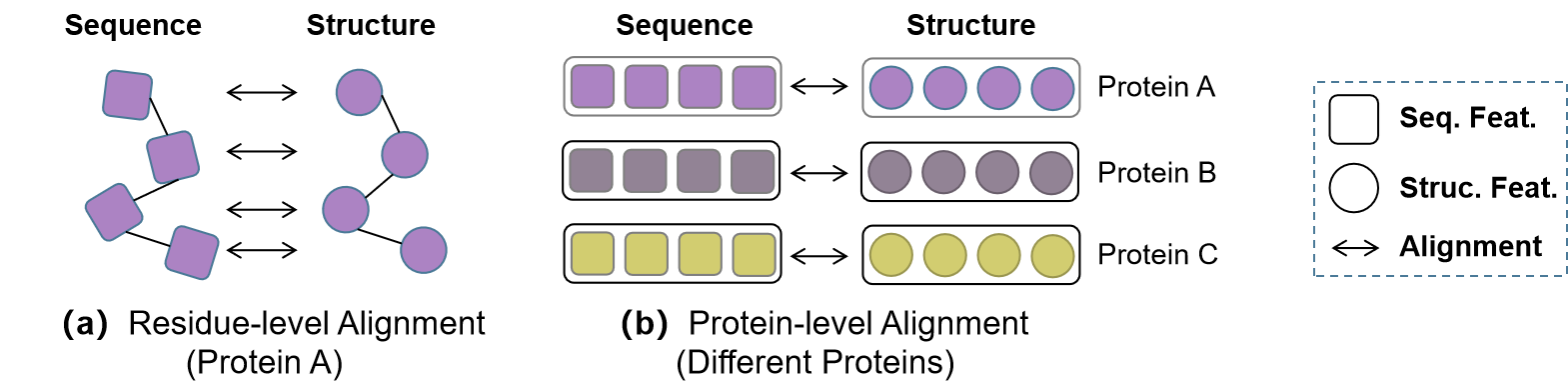}
   \vspace{-1em}
   \caption{Various alignment levels. (a) Residue-level alignment entails comparing each pair of structure-sequence features residue by residue. (b) Protein-level alignment involves comparing each pair of structure-sequence features protein by protein. The features of each protein are amalgamated from all the residue features contained within it. Identical colors indicate a sequence-structure pair originating from the same protein.}
   \label{:align_level}
   \vspace{-1em}
\end{figure*}

\textbf{Contrastive Alignment Level.}
As shown in Figure \ref{:align_level}, we propose to compute the alignment loss at both the residue and protein levels, respectively. 
For residue-level alignment, we compare the encoded sequence and structure features residue by residue. In contrast, for protein-level alignment, we compare the encoded features at a coarser, fine-grained level. 
Intuitively, alignment at different fine-grained levels results in different capabilities. Finer-grained comparisons, such as residue-level alignment, necessitate more complex computations but may yield better performance. Conversely, coarse-grained comparison alignments, such as protein-level alignment, may exhibit inferior performance but are worth considering due to their lighter computational load. Refer to our experimental section for a detailed analysis.
Regardless of the level used for calculation, the samples in the mini-batch will be randomly shuffled, and each pairing will be different, thereby implicitly serving as a form of data augmentation.

\begin{figure*}[htp]
  \centering
   \includegraphics[width=0.82\linewidth]{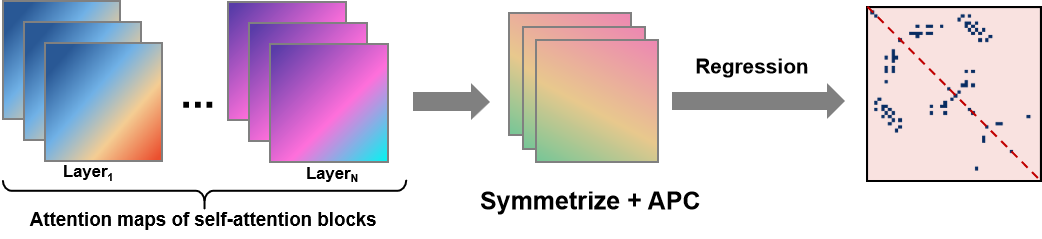}
   \vspace{-1em}
   \caption{Pipeline for reconstructing the contact map based on C$_\beta$ atoms with length \(L\): First, attention maps are extracted from each layer of the self-attention blocks. These maps undergo symmetrization and average product correction (APC) along the amino acid dimensions to produce an \(L \times L\) coupling matrix. This matrix forms the basis for the final contact map predictions, which are refined using a regression layer.}
   \label{:contact}
   \vspace{-1em}
\end{figure*}

\textbf{Structural Reconstruction Constraint.}
As previously mentioned, the GVP layers encode the core structural features using N, C\text{$_\alpha$}, and O atoms (These three types of atoms are called backbone atoms). To further enhance the structural constraints, we introduce a self-supervised structure reconstruction task. While predicting the coordinates of virtual C\text{$_\beta$} atoms directly from the structural features of N, C\text{$_\alpha$}, and O atoms would be the most straightforward approach, it proves challenging to achieve accurate predictions in practice. Therefore, we simplify this task by transforming the coordinate prediction into a contact map prediction, illustrated in the Figure~\ref{:contact}. Specifically, we utilize the intermediate attention maps generated by the self-attention blocks depicted in Figure \ref{fig:overall}(a) to predict the contact maps related to C\text{$_\beta$} \cite{rao2020transformer}.
In addition to serving as a constraint during training to preserve structural information, the generation of contact maps can also serve as a metric for internally evaluating the performance of the pretrained model.

\textbf{Virtual \text{C$_{\beta}$} Atom Generation.}
The virtual C$_\beta$ atoms are derived coordinates that may not physically exist in every residue. However, their positions can be inferred from the spatial relationships among protein backbone atoms: N, C$\alpha$, and O atoms. The positions are calculated as follows:
\begin{equation}
\left\{
    \begin{aligned}
        \delta_0 & = \varrho(C_{\alpha}) - \varrho(N), \\
        \delta_1 & = \varrho(C) - \varrho(C_{\alpha}), \\
        \delta_2 & = \delta_0 \otimes \delta_1,\\
        \varrho(C_{\beta}) & =  \epsilon_0 \cdot \delta_0 + \epsilon_1 \cdot \delta_1 +  \epsilon_2 \cdot \delta_2 + \varrho(C_{\alpha}),
    \end{aligned}
    \right.
\end{equation}
where $\otimes$ denotes the cross product of vectors and $\varrho(*)$ denotes the coordinates of the corresponding atom types. And $\epsilon_0=0.5680$, $\epsilon_1=-0.5407$ and $\epsilon_2=-0.5827$ are constant coefficients.

\textbf{Self-supervised Contact Map Prediction.}
The pipeline diagram of the contact map predictor is depicted in Figure~\ref{:contact}. Self-attention maps extracted from the self-attention blocks undergo symmetrization and average product correction (APC) to generate the final contact maps. The self-supervised structural loss is formulated to constrain the structural feature representation using cross-entropy:
\begin{equation}
\begin{split}
    \mathcal{L}_{contact} & = \text{CrossEntropy}(\text{Sigmoid}(\text{M}_{pred}),\text{M}_{ref})\\
    & = - \sum_{i=1} \text{M}_{ref}^{(i)} \cdot \text{log}(\text{M}_{pred}^{(i)}),
\end{split}
\end{equation}
where $\text{M}_{ref}$ denotes the ground-truth contact maps (binary values 0 or 1), and $\text{M}_{pred}$ denotes the predicted values distributed between 0 and 1.

The pretraining objective involves jointly optimizing the contrastive alignment loss $\mathcal{L}_{\text{align}}$ and the contact map loss $\mathcal{L}_{\text{contact}}$, with weighted values $\lambda_1$ and $\lambda_2$, as expressed by::
\begin{equation}
   \mathcal{L}_{\text{pretrain}} = \lambda_1 \cdot \mathcal{L}_{\text{align}} + \lambda_2 \cdot \mathcal{L}_{\text{contact}}.
\end{equation}

\subsection{Inference Phase}
As depicted in Figure~\ref{fig:overall}(b), the architecture of the inference phase utilizes only the pretrained language-enhanced structure model, rendering other flows unnecessary during this stage.
Evaluating a pretrained protein structure model within a novel training framework poses significant challenges. To address these challenges, we have developed a comprehensive evaluation system that includes multiple validation tasks, showcasing the model's representation learning, alignment capability, and generalization ability. Based on the necessity for fine-tuning, we categorize the validation tasks into internal and external/downstream tasks.

\textbf{External Evaluation Tasks.}
External tasks require fine-tuning and are focused on downstream applications. We introduce the protein sequence design task (also known as protein inverse folding), which involves predicting protein sequences based on corresponding protein backbone atomic coordinates. Key metrics for this task include perplexity and accuracy of sequence recovery. Additionally, we incorporate protein functional recognition tasks to validate the robust representation capabilities acquired during training.

\textbf{Internal Evaluation Tasks.}
Internal tasks do not require fine-tuning and include contact map prediction and self-similarity evaluation. The contact map prediction task uses the top-L long-range precision (P@L) metric to evaluate the quality of the predicted contact maps. The self-similarity evaluation task assesses the alignment between the language model and the structure model using accuracy and KL divergence metrics.

\section{Experiments}

\vspace{-2em}
\begin{table*}[htp]
\centering
\scalebox{0.75}{
\begin{tabular}{llcccc} 
    \toprule

    \multirow{2}{*}{\textbf{Input}} & \multirow{2}{*}{\textbf{Methods}} & \multicolumn{3}{c}{\textbf{Gene Ontology}} &  \textbf{Enzyme Commission} \\ 

    \cmidrule(lr){3-5} 
    
     &  & \textbf{BP} & \textbf{MF} & \textbf{CC} & \\ 
    
    \midrule

   \multirow{4}{*}{1D} & CNN & 0.244 & 0.354 & 0.287 & 0.545\\ 
      & ResNet & 0.280 & 0.405 & 0.304 & 0.605\\
      & LSTM & 0.225 & 0.321 & 0.283 & 0.425\\
      & Transformer & 0.264 & 0.211 & 0.405 & 0.238\\

      \midrule

      \multirow{3}{*}{1D} & GCN & 0.252 & 0.195 & 0.329 & 0.320\\ 
      & GAT & 0.284 & 0.317 & 0.385 & 0.368\\
      & 3D CNN & 0.240 & 0.147 & 0.305 & 0.077\\
  
      \midrule

      \multirow{5}{*}{3D+1D} & GraphQA & 0.308 & 0.329 & 0.413 & 0.509\\
        & GVP~\cite{BowenJing2020LearningFP} & 0.326 & 0.426 & 0.420 & 0.489 \\
        & IEConv (residue level)~\cite{PedroHermosilla2020IntrinsicExtrinsicCA} & 0.421 & 0.624 & 0.431 & -\\
        & GearNet~\cite{ZuobaiZhang2022ProteinRL} & 0.356 & 0.503 & 0.414 & 0.730\\
        & CDConv~\cite{fan2022continuous} & 0.453 & 0.654 & 0.479 &  0.820\\
    
      \midrule

       3D & Ours & \textbf{0.459} & \textbf{0.663} & \textbf{0.491} & \textbf{0.828}\\
    
	 \bottomrule
  
\end{tabular}
}
\caption{\label{tab:go} F$_{max}$ of gene ontology term prediction and enzyme commission prediction. }
\vspace{-2em}
\end{table*}

\subsection{Settings}

\textbf{Pretraining Datasets.} We utilize a larger-scale protein dataset, PDB, comprising sequence-structure pairs, as the pretraining dataset. To prevent label leakage, we exclude all existing data also appearing in evaluation datasets.
To augment the datasets, we use the AlphaFoldDB for pretraining. This database contains the protein structures predicted by the AlphaFold2 model.

\textbf{Evaluation Datasets.} The CATH dataset is widely employed, featuring training, validation, and test splits consisting of 18204, 608, and 1120 protein data samples. In downstream protein design tasks, the training set is utilized for finetuning, and the test set is used for evaluation. We also report results on Ts50 \& Ts500 \cite{li2014direct}. 
Furthermore, the trRosetta set is utilized for contact map prediction, comprising around 15000 instances. The CASP14 set, renowned for AlphaFold2, though modest in number, closely resembles the practical environment of blind tests and competitions. To summarize, the trRosetta set, CATH testing set, Ts50/Ts500, and CASP14 set are all utilized for internal evaluation.

\textbf{Implementation Details.} 
During pretraining, AdamW optimizer with a batch size of 8 and an initial learning rate of 1e-3 is used. We employ the ESM-2 base version as the default teacher protein language model, with fixed parameters in the training pipeline. The GVP module comprises 4 layers with a dropout of 0.1, top-k neighbors of 30, a node hidden dimension of scalar features of 1024, and a node hidden dimension of vector features of 256.
The self-attention block following the GVP includes 4 self-attention layers, 8 attention heads, an embedded dimension of 512, and an attention dropout of 0.1. It's noteworthy that we pretrain the model separately for residue-level alignment and protein-level alignment.
2 NVIDIA GPU A100 80GB were used.

\subsection{Evaluation on Protein Function Prediction Tasks}

\begin{table*}[htp]
\vspace{-2em}
\centering
\scalebox{0.8}{
\begin{tabular}{llcccccc} 
    \toprule

    \multirow{2}{*}{\textbf{\#}} & \multirow{2}{*}{\textbf{Models}} & \multicolumn{3}{c}{\textbf{Perplexity}} & \multicolumn{3}{c}{\textbf{Recovery (\%)}} \\ 

    \cmidrule(lr){3-5}\cmidrule(lr){6-8}
    
    &  & \textbf{CATH} & \textbf{Ts50} & \textbf{Ts500} & \textbf{CATH} & \textbf{Ts50} & \textbf{Ts500} \\ 
    \midrule
 
    \multirow{9}{*}{1} 
    & Natural frequencies \cite{hsu2022learning}  & 17.97 & - & -  & 9.5 & - & -\\
    & SPIN2 \cite{o2018spin2}  & - & - & -  &- & 33.6 &36.6 \\
    & Structured Transformer \cite{ingraham2019generative}  & 6.85 & 5.60 & 5.16  & 36.4 & 42.40 &44.66 \\
    & Structured GNN \cite{jing2020learning}   & 6.55 & 5.40 & 4.98  & 37.3 & 43.89 & 45.69\\
    & GVP-Transformer \cite{hsu2022learning}   & 6.44 & - & - & 38.3 & - &-\\
    & AlphaDesign \cite{gao2022alphadesign}  & 6.30 & 5.25 & 4.93  & 41.31 & 48.36 & 49.23\\
    & GVP-GNN-large \cite{jing2020learning}  & 6.17 & - & - & 39.2 & - &-\\
    & GVP-GNN \cite{jing2020learning}   & 5.29 & 4.71 & 4.20  & 40.2 & 44.14 & 49.14 \\
    & ProteinMPNN \cite{dauparas2022robust}   & 4.61 & 3.93 & 3.53  & 45.96 & 54.43 & 58.08\\
    
    \midrule
    
    \multirow{3}{*}{2}
    
    & $\text{Design}$ (\textit{w/o} \text{Pretraining})	& 6.27	& 5.05 & 4.87 & 39.62 & 49.21 & 50.01 \\
    & $\text{Design}_{p}$ (\textit{w/} \text{Pretraining})  & \underline{4.51} & \underline{3.82} & \underline{3.35} & \underline{50.1} & \underline{55.7} & \underline{59.5}\\ 
    
    & $\text{Design}_{r}$ (\textit{w/} \text{Pretraining})  & \textbf{4.48} & \textbf{3.76}  & \textbf{3.28} & \textbf{50.8} & \textbf{55.8} & \textbf{60.3}\\
	 \bottomrule
  
\end{tabular}
}
\caption{\label{tab:design} Comparison among our protein design models (\#2) and baselines (\#1). The best results are \textbf{bolded}, followed by \underline{underlined}. Design$_p$: Protein-level pretrained model; Design$_r$: Residue-level pretrained model.}
\vspace{-2em}
\end{table*}

Our method shows superior performance in gene ontology (GO) term prediction and enzyme commission (EC) number prediction tasks, surpassing existing 1D-only, 3D-only, and (3+1)D approaches. We used the F$_{max}$ accuracy metric for evaluation. These findings, detailed in Table~\ref{tab:go}, highlight the effectiveness of our cross-modal contrastive learning approach.

\begin{figure}
    \centering
    \includegraphics[width=0.9\linewidth]{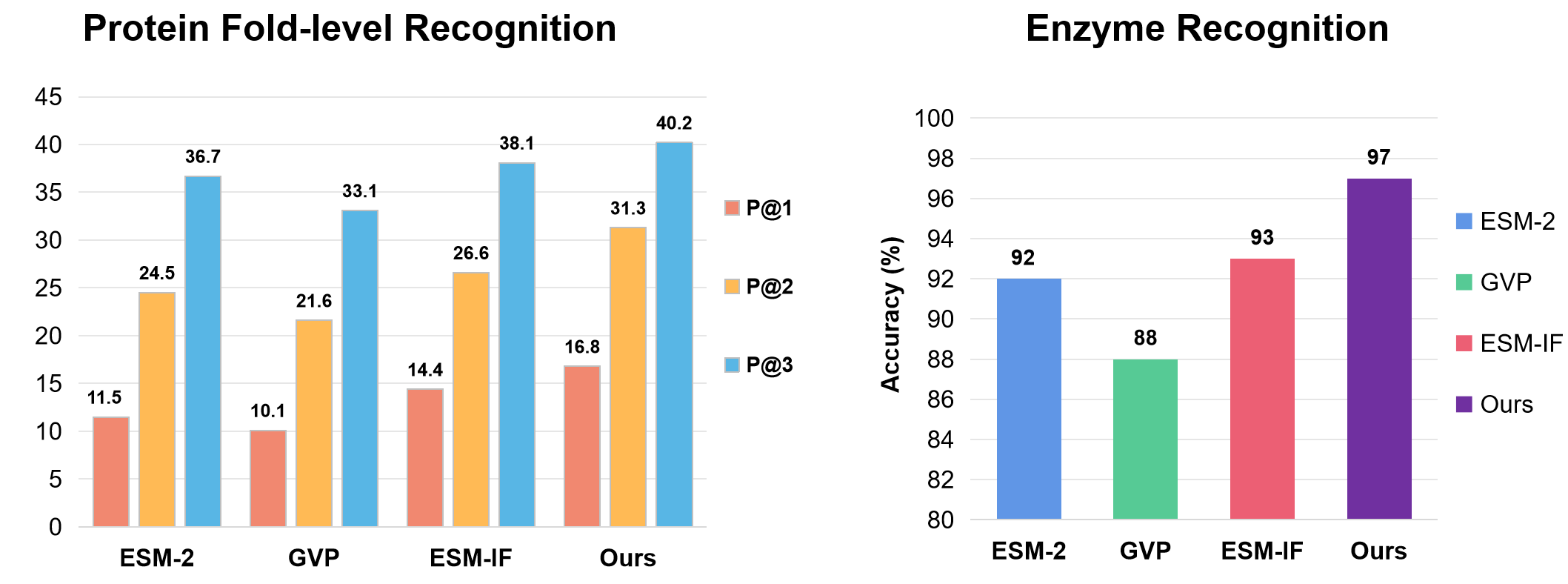}
    \vspace{-1em}
    \caption{Protein functional prediction tasks. (a) Comparing the fold-level predictions. P@k denotes the top-k precision. (b) Performances for enzyme recognition task.}
    \label{fig:func}
    \vspace{-2em}
\end{figure}

The IEConv method, featuring both atom-level (Hermosilla et al., 2021) and amino-acid-level (Hermosilla \& Ropinski, 2022) variants, served as a key benchmark in our comparisons. The atom-level variant, denoted as "3D+Topo," utilizes 3D coordinates and the topological structure of bonds between atoms. Our approach outperformed this and other existing methods significantly.
Following the work of Hermosilla et al. (2021), Hermosilla \& Ropinski (2022), and Zhang et al. (2022), we assessed our method across three GO term prediction sub-tasks: biological process (BP), molecular function (MF), and cellular component (CC). Both GO term and EC number predictions are framed as multi-label classification tasks.
In essence, our approach integrates 1D and 3D data using cross-modal contrastive learning, yielding robust representations and achieving higher accuracy in GO and EC prediction tasks.

\subsection{Evaluation on Protein Inverse Folding Tasks}

Computational protein design, also known as \textit{protein inverse folding},  aims to deduce amino acid sequences given the corresponding atomic coordinates of protein backbones. The protein design model serves as a key component for downstream evaluation. It directly leverages the pretrained structure model as the backbone, coupled with a non-autoregressive decoder featuring linear MLPs. The CATH testing set and Ts50/Ts500 are employed to evaluate the primary results.
Table~\ref{tab:design} presents a selection of several prominent baselines across different types. Notably, ProteinMPNN \cite{dauparas2022robust} and GVP-Transformer \cite{hsu2022learning} exhibit advanced performance, particularly excelling in sequence recovery and perplexity. The lightweight GVP-GNN \cite{jing2020learning} also demonstrates competitiveness, showcasing relatively strong performance and speed.
Our protein design models outperform the baselines. Notably, the inclusion of a non-autoregressive decoder in our module contributes to faster sampling speeds. 
Furthermore, regarding different levels (Design$_p$ vs. Design$_r$), while there is no significant disparity between the residue-level and protein-level pretrained modules in downstream tasks, Design$_r$ slightly outperforms Design$_p$ overall. This observation suggests that any small gaps between the two levels of pretrained models are further diminished during the fine-tuning process.
Additionally, within Group 2 of Table~\ref{tab:design}, we compared the pretrained model with a non-trained model serving as the backbone. Notably, the pretrained model significantly enhances performance in terms of perplexity and recovery, underscoring the stronger generalization ability of the pretrained structure model for downstream tasks.

\subsection{Evaluation on Protein Fold-level Classification}
Furthermore, we undertake predictions on challenging fold types based on the Fold dataset~\cite{hermosilla2020intrinsic}, which essentially involves a less data-intensive multi-task enzyme function prediction.
We gather approximately 700 enzymes with experimentally determined structure-sequence pairs across 10 folds from RCSB for multi-class functional prediction tasks. This involves precisely predicting the fold level to which an enzyme belongs.
`Fold' here refers to the 3D arrangement of secondary structural elements (such as alpha helices and beta sheets) that characterize a particular protein or group of proteins. Proteins with similar folds typically share significant structural similarities, even if their sequences and functions differ. Fold classification can offer insights into evolutionary relationships among proteins. 
The objective of this setup is to predict, within enzymes with mixed folds, the specific fold to which an enzyme belongs. As illustrated in Figure~\ref{fig:func}(a), we choose ESM-IF and GVP as baseline structure-to-sequence models, as they also utilize N, C$_\alpha$, and C as standard inputs. These comparisons validate the superior structural representation capabilities by a large margin (Ours: 16.8\%P@1, 31.3\%P@2, 40.2\%P@3). We employ ESM-2 as a language modality input for comparison to confirm its sequence-only representation capabilities. Although ESM-2 (11.5\%P@1, 24.5\%P@2, 36.5\%P@3) slightly outperforms the earlier GVP, it significantly lags behind the performance of ours, demonstrating enhanced generalization ability due to the incorporation of context information.

\subsection{Evaluation on Functional Enzyme Recognition}
To explore enzyme recognition through binary classification, we leverage a novel dataset that extends the Fold dataset. Each fold in this dataset is meticulously crafted to include an equal number of enzymes, balanced with non-enzyme negative samples. Given the binary nature of the classification task, positive and negative samples are aggregated across folds and then randomly divided, with 80\% allocated to the training set and 20\% to the evaluation set.
Our experimental findings, as illustrated in Figure~\ref{fig:func}(b), reveal that our model, augmented with prior language enhancements, achieves the highest performance, boasting an average accuracy of 97\%. Among the baseline models, ESM-IF and ESM2 yield comparable accuracies of 93\% and 92\%, respectively, despite operating on different modalities (sequence vs. structure). Notably, our model outperforms ESM-IF, a benchmark recognized for its exceptional representation and similarity in modalities, indicating superior generalization capabilities.

\begin{figure}
    \centering
    \includegraphics[width=0.83\linewidth]{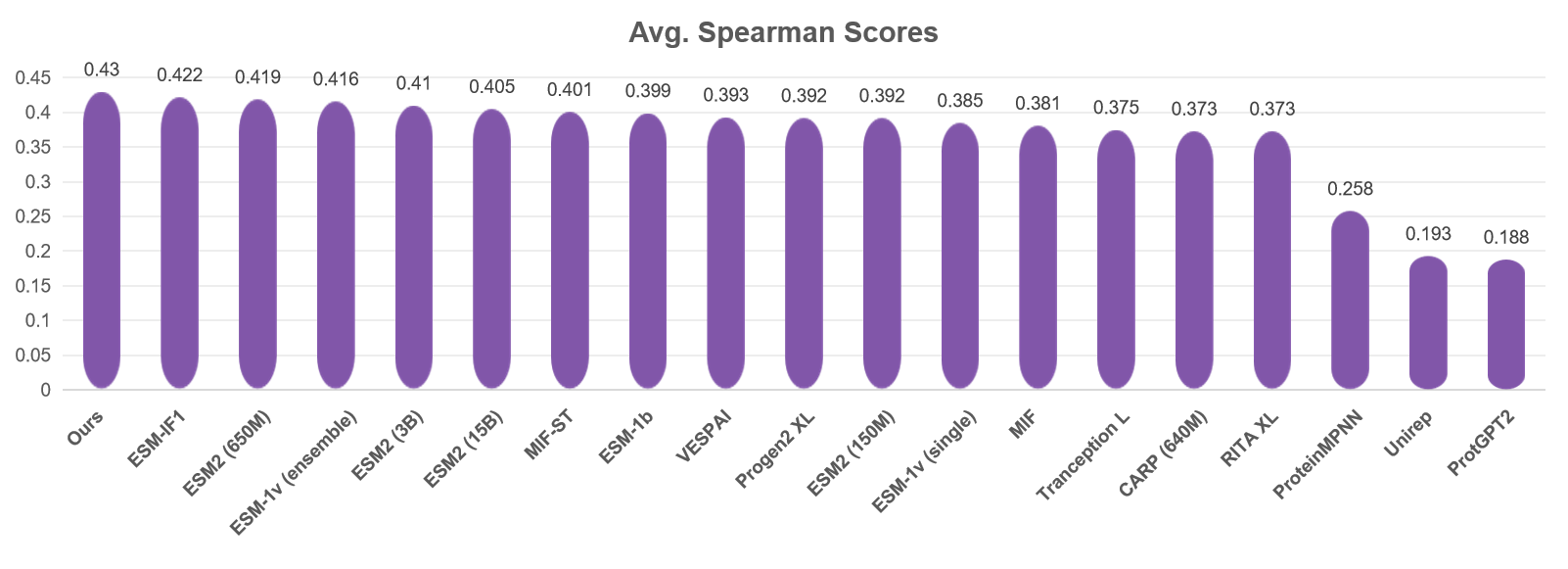}
    \vspace{-1em}
    \caption{Spearman’s rank correlation on ProteinGym set.}
    \label{fig:proteingym_scores}
\end{figure}

\subsection{Zero-shot Learning for Protein Fitness Prediction}
Additionally, we introduce a zero-shot learning approach for fitness prediction, which enables the direct validation of the model's representational stability in a non-parametric manner. To ensure a thorough and unbiased comparison, we selected prominent protein language models and inverse folding models as benchmarks, as depicted in Figure~\ref{fig:proteingym_scores}. All baseline models utilize zero-shot learning for evaluating mutation effects through the ProteinGym dataset~\cite{notin2023proteingym}, which encompasses millions of mutations.
Our proposed method yields an average $\rho$ of 43.0\%, consistently achieving the top matching rank. This suggests that leveraging prior language knowledge significantly contributes to enhanced overall performance in mutation prediction. In contrast, the optimal baseline model (ESM-IF) scores 42.2\%. The superiority in performance of our model can be attributed to its dual advantage, encompassing both contextual and structural transfer learning within the proposed training paradigm.

\section{Ablation Studies and Analysis}

\vspace{-2em}
\begin{table*}[htp]
\centering
\scalebox{0.63}{
\begin{tabular}{ccccccccccccccccccc} 
    \toprule
    
    \textbf{Group} & \textbf{Level} & \multicolumn{3}{c}{\textbf{CATH Test Set}} & \multicolumn{3}{c}{\textbf{trRosetta Set}} & \multicolumn{3}{c}{\textbf{Ts500 Set}} & \multicolumn{3}{c}{\textbf{Ts50 Set}} & \multicolumn{3}{c}{\textbf{CASP14 Set}} \\ 

    \midrule
    & & \textbf{P@L} & \textbf{P@L2}& \textbf{P@L5}& \textbf{P@L} & \textbf{P@L2}& \textbf{P@L5}& \textbf{P@L} & \textbf{P@L2}& \textbf{P@L5}& \textbf{P@L} & \textbf{P@L2}& \textbf{P@L5} & \textbf{P@L} & \textbf{P@L2}& \textbf{P@L5}\\
        
    \cmidrule(lr){3-17}
        
    \multirow{2}{*}{1}        
    & Residue & 78.05& 90.97 & 96.12 & 87.69	 & 94.94	 & 96.89	 & 90.31	 & 96.67	 & 98.10	 & 91.36	 & 98.66	 & 100.00 & 74.9 & 91.49 & 95.34 \\
     & Protein & 72.21	& 88.03	& 98.31	& 81.30	& 92.42	& 96.18	& 83.78	& 94.14	& 97.44	& 85.18	& 96.32	& 99.53 & 68.24 & 87.23 & 93.96 \\

    \midrule

     & & \textbf{acc$_1$} & \textbf{acc$_2$}& \textbf{KL}& \textbf{acc$_1$} & \textbf{acc$_2$}& \textbf{KL}& \textbf{acc$_1$} & \textbf{acc$_2$}& \textbf{KL}& \textbf{acc$_1$} & \textbf{acc$_2$}& \textbf{KL}& \textbf{acc$_1$} & \textbf{acc$_2$}& \textbf{KL}\\

     \cmidrule(lr){3-17}
     
    \multirow{2}{*}{2} 
    & Residue & 87.71 & 87.65 & 0.004 & 96.13 & 96.23 & 0.0001 & 94.69 & 95.02 & 0.0001 & 98.29	 & 98.41 & 0.0001 & 30.87 & 30.45 & 0.0001  \\
     & Protein & 100.00 & 100.00 & 0.00 & 100.00 & 100.00 & 0.00 & 100.00 & 100.00 & 0.00 & 100.00 & 100.00 & 0.00 & 100.00 & 100.00 & 0.00 \\
	 \bottomrule
\end{tabular}
}
\caption{\label{tab:1} 
Internal evaluation across test sets. Group 1 showcases the contact map predictions, measured by P@* scores. Group 2 focuses on retrieval alignment evaluations, quantified by alignment accuracy and KL distance. The acc$_1$ and acc$_2$ metrics denote the accuracy of structure-to-sequence and sequence-to-structure alignment.}
\vspace{-2.5em}
\end{table*}

\subsection{Internal Contact Map Generation}
Contact map predictions serve as a significant indicator of structural representation capabilities. Due to our pretraining mechanism, the contact map predictor can generate accurate contact maps directly without the need for fine-tuning.
Group 1 of Table~\ref{tab:1} showcases the contact map prediction scores, evaluated across the CATH, trRosetta, Ts50/Ts500, and CASP14 test sets. Both the residue-level and protein-level pretrained models demonstrate high P@L accuracy in predicting contact maps across all datasets, indicating that the pretrained structure module has acquired rich structural representations. Notably, the residue-level evaluation exhibits superior performance within Group 1, likely attributable to its finer granularity.
Expanding on this analysis, the ability of our pretrained models to generate precise contact maps across diverse datasets underscores their robustness and generalization potential in capturing intricate structural details. Such proficiency in contact map prediction signifies the effectiveness of our pretraining approach in enhancing structural representation learning.

\subsection{Retrieval Alignment Evaluation}
In the proposed pretraining framework, we operate under the strong assumption that protein language models and protein structure models are equally proficient in representing features, albeit through different modalities. Hence, we advocate for quantifying the sequence-structure retrieval power to gauge the alignment prowess of the pretrained model. 
Reflecting on the contrastive alignment loss employed during pretraining, it becomes evident that the loss encompasses both structure-to-sequence and sequence-to-structure alignment calculations. The intermediate state score computations offer a direct means to evaluate the multi-modality alignment level.
The retrieval alignment evaluation on the CATH, trRosetta, Ts50/Ts500, and CASP14 test sets is presented in Group 2 of Table~\ref{tab:1}. Additionally, we provide residue-level and protein-level results for comprehensive analysis. It's noteworthy that the protein-level pretrained model exhibits a higher ease in aligning sequences and structures, evident through higher accuracy and lower KL distance, which aligns well with our intuition. 
Overall, the pretrained structure module demonstrates a robust alignment level, underscoring the effectiveness of the proposed framework.

\begin{figure}
    \centering
    \includegraphics[width=0.4\linewidth]{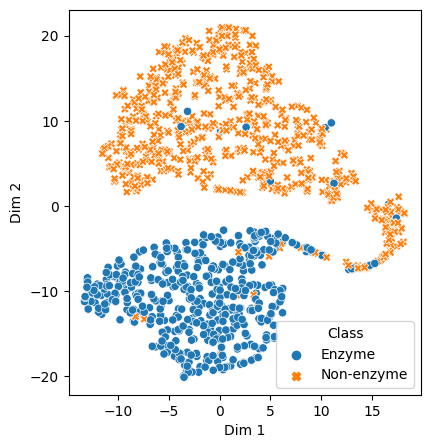}
    \vspace{-1.5em}
    \caption{Visualization using t-SNE to demonstrate enzyme recognition.}
    \label{fig:tsne_fig}
    \vspace{-2.5em}
\end{figure}

\subsection{Classification Visualization}
To visually represent the efficacy of enzyme classification, we advocate the utilization of t-distributed Stochastic Neighbor Embedding (t-SNE) as a means to illustrate the clustering patterns of enzymes, as depicted in Figure~\ref{fig:tsne_fig}. Employing this visualization technique provides insights into the distribution of enzyme data points in a lower-dimensional space.
In the context of enzyme recognition, our proposed structural model manifests discernible classification boundaries within the t-SNE plot. This observation underscores the robustness of our model's feature representation capabilities. The distinct clustering of enzymes reaffirms the model's ability to delineate between enzyme and non-enzyme samples effectively, further validating suitability for the task of functional recognition.

\section{Related Work}
\textbf{Protein Language Models.}
Protein language modeling has emerged as a promising avenue for unsupervised learning of protein primary structures~\cite{zheng2024metaenzyme,zheng2023mmdesign,hu2023learning}. Models such as UniRep~\cite{EthanCAlley2019UnifiedRP} utilize LSTM or its variants to capture sequence representations and long-range dependencies. TAPE~\cite{RoshanRao2019EvaluatingPT} benchmarks a range of protein models across various tasks, affirming the effectiveness of self-supervised pretraining methods. Many efforts have focused on enhancing model scale and architecture to capture richer protein semantics. For instance, ESM-1b employs a Transformer architecture and masked language modeling strategy to learn robust representations from a large-scale dataset. Subsequently, ESM-2~\cite{ZemingLin2022LanguageMO} extends ESM-1b with larger-scale parameters (15 billion), achieving superior results compared to smaller ESM models.

\textbf{Protein Structure Models.}
Given that a protein's function is often determined by its structure. The advancements in protein structure prediction methods have led to initiatives like AlphaFoldDB, providing over 200 million protein structure predictions to accelerate scientific research. Building upon this progress, various protein structure models have emerged, aiming to encode spatial information using convolutional neural networks (CNNs) or graph neural networks (GNNs). 
Among these methods, IEConv~\cite{PedroHermosilla2020IntrinsicExtrinsicCA} introduces a convolution operator to capture all relevant structural levels of a protein. GearNet~\cite{ZuobaiZhang2022ProteinRL} encodes the spatial information by adding different types of sequential or structural edges and then performed relational message passing on protein residue graphs. GVP-GNN~\cite{BowenJing2020LearningFP} designed the geometric vector perceptrons (GVP) to learn both scalar and vector features in an equivariant and invariant manner, while \cite{YuzhiGuo2022SelfsupervisedPF} adopts SE(3)-invariant features as model inputs and reconstruct gradients over 3D coordinates to avoid the complexity of SE(3)-equivariant models.
 
\section{Conclusions and Limitations}
We propose leveraging pretrained protein language model to train protein structure models using cross-modal contrastive learning . Our approach demonstrates superior performances in various evaluation tasks. However, challenges remain, including the scope of language model transfer, data efficiency, generalization, computational resources, and evaluation metrics. Addressing these limitations will be crucial for advancing the utility of pretrained protein language models in protein structure prediction and related applications.

\section*{Acknowledgments}
This work was supported by National Science and Technology Major Project (No. 2022ZD0115101), National Natural Science Foundation of China Project (No. U21A20427), Project (No. WU2022A009) from the Center of Synthetic Biology and Integrated Bioengineering of Westlake University and Integrated Bioengineering of Westlake University and Project (No. WU2023C019) from the Westlake University Industries of the Future Research Funding.

\bibliographystyle{splncs04}
\bibliography{my_citation}

\end{document}